\documentclass[showpacs,preprint,amssymb]{revtex4}
\usepackage{graphicx}
\usepackage{dcolumn}
\usepackage{bm}
\def\be{\begin{equation}} 
\def\ee{\end{equation}}
\def\bea{\begin{eqnarray}} 
\def\eea{\end{eqnarray}}
\def\line{\hbox to \hsize}    
\def\frac #1#2{{#1\over #2}}

\def\psid{\psi^{\dagger}}

\def \ket #1{{\vert #1\rangle}}

\def\eval #1#2#3{{\langle#1\vert#2\vert#3\rangle}} 
\def\vev #1{{\langle #1\rangle}}
\def\1{\mbox{\bf 1}}

\begin{document}

\title{Josephson Currents in  Quantum Hall Devices}

\author{ MICHAEL STONE}

\affiliation{University of Illinois, Department of Physics\\ 1110 W. Green St.\\
Urbana, IL 61801 USA\\E-mail: m-stone5@illinois.edu}   

\author{YIRUO LIN}

\affiliation{University of Illinois, Department of Physics\\ 1110 W. Green St.\\
Urbana, IL 61801 USA\\E-mail: yiruolin@illinois.edu}

\begin{abstract}  
We consider a simple model for an  SNS Josephson junction in which the ``normal metal''  is a section of a filling-factor $\nu=2$ integer  quantum-Hall edge.  We provide analytic expressions for the current/phase relations to all orders in the coupling between the superconductor and the quantum Hall edge modes, and for all temperatures.  Our conclusions are consistent with the earlier perturbative study by Ma and Zyuzin [Europhysics Letters {\bf 21} 941-945 (1993)]:  The  Josephson current is independent of the distance between the superconducting leads, and the upper bound on the maximum Josephson current is inversely proportional to the perimeter of the Hall device.

\end{abstract}


\maketitle

\section{Introduction}

The zero-voltage Josephson  current  in a supercondictor/normal-metal/superconductor (SNS) junction \cite{waldramSNS}
arises from  Andreev scattering  \cite{andreev} at  the  SN and  NS interfaces.  In the ideal case,  an   electron incident on one superconductor from the normal metal will  be reflected back into the normal metal as a hole,  and this hole, on striking the second superconductor, will  be reflected back towards the first  superconductor as  an   electron.  When the relative phase of the order parameters is  such that constructive interference occurs,  the   back-and-forth  process    continues {\it ad infinitum\/}  and transfers  two electrons from superconductor to superconductor in each   cycle \cite{maslov,fazio1,takane,fazio2,afleck}. A round trip  takes time $2W/v_F$, where $v_F$ is the Fermi velocity and $W$ the separation   between the superconductors. The current will therefore be $ev_F/W$ for each  open transverse channel.  In practice, the probability of Andreev reflection is less than unity \cite{griffin,blonder} and the motion in the metal may  be  diffusive, but $ev_F/W$ per channel  remains   an upper bound on the critical current.  

An interesting 
question arises as to what happens when the ``normal'' metal consists of the chiral fermions at the edge of  a quantum Hall (QH) bar
 \cite{wen-edge}.  
  In this case the holes move in the {\it  same\/}  direction as the electrons, so  conventional Andreev retro-reflection is impossible. A two-electron charge transfer requires  a (phase coherent) passage  around the entire perimeter of the Hall bar, and  this lengthy excursion  suggests that the small ``$W$'' of the conventional junction be replaced by the much larger perimeter $L$ of the Hall bar.   A perturbative study of a  S-QH-S system  in \cite{ma}  supports this conclusion  and estimates    that the maximum Josephson current will be very small --- in the order of 1 nA for mm scale  devices.  In view of ongoing experiments on quantum-Hall Josephson junctions, however, it seems  worth revisiting the problem to see if devices might  be engineered to provide larger critical currents.

In this paper we introduce a model of an S-QH-S junction that  is simple enough that it can be studied non-perturbatively. We obtain analytic expressions for the Josephson current/phase relation to all orders in the S-QH coupling, and at all temperatures.  Despite our greater control over the model, the   key conclusions of the perturbative studies in \cite{ma} (see also \cite{zyuzin}) are unchanged: at filling fraction $\nu=2$  an   upper bound for the critical Josephson current is given by  $2ev_d/L$ where $v_d$ is the edge-mode drift velocity and $L$ is the  perimeter of the Hall device. Further, the temperature scale at which the Jospehson current is washed out by thermal effects is set by the edge-mode level spacing $E_{n+1}-E_n= 2\pi \hbar v_d/L$.  Thus, if we wish to see Josephson-junction physics in quantum Hall devices, we should construct the junctions by  coupling  superconducting probes  to meso-scale Hall-dots.

 In section two we introduce the model and solve the associated Bogoliubov-de Gennes equation. In section three we introduce an analytic regularization scheme  to handle the otherwise ill defined sums  that appear in the  current/phase relation. In section four we demonstrate  that our  regularization  scheme is  consistent with conventional perturbation theory at both zero and non-zero temperature. We finish with a brief discussion of effects that we have not taken into account, and  that may or may not be significant.

\section{The model}

\begin{figure}
\includegraphics[width=3.5in]{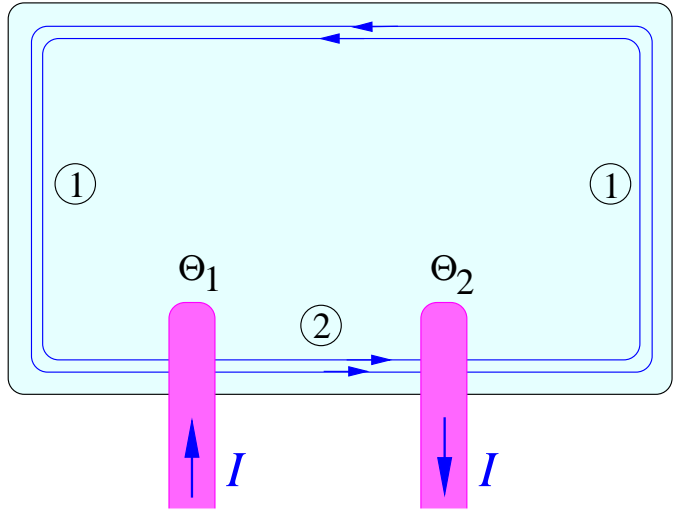}
\caption{\sl A Hall bar with superconducting probes passing a  current $I$ through the edge modes. The circled numbers label the regions (1) ``outside the leads,'' and (2) ``between the leads.''}
\label{FIG:Hallbar}
\end{figure}

\begin{figure}
\includegraphics[width=3.5in]{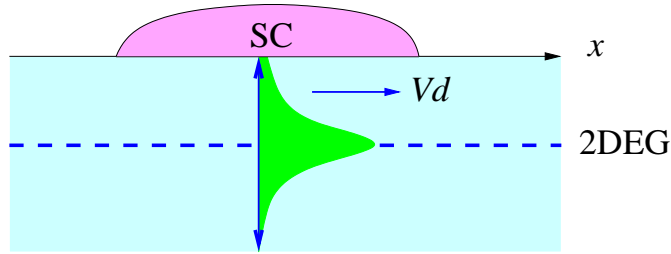}
\caption{\sl  The wavefunction for an electron in a 2DEG is  confined in the vertical direction,  but there is some amplitude for the  vertically oscillating  electron to touch  the superconductor. As a slowly-drifting   Landau-level  wavepacket passes under  the superconducting lead, there will be   many opportunities for  Andreev reflection to  turn the   electron into a hole.}
\label{FIG:Edge}
\end{figure}

We  consider a  $\nu=2$ quantum-Hall edge  (two spins therefore) in interaction with   superconducting (SC) leads (figure \ref{FIG:Hallbar}). We  model the system   by a linear-dispersion   edge-mode hamiltonian 
\bea
H=\oint  \left\{-iv_d\psid_\uparrow(\partial_x-ieA)\psi_\uparrow  -iv_d\psid_\downarrow(\partial_x-ieA)\psi_\downarrow \right.\quad&&\nonumber\\
\left. + |\Delta(x)|e^{i\theta(x)}\psid_\uparrow\psid_\downarrow  +|\Delta(x)|e^{-i\theta(x)}\psi_\downarrow \psi_\uparrow \right\}dx.&&
\eea
Here $v_d$ is the edge-mode drift velocity that is proportional to the gradient of the confining potential.
 The terms with $\Delta(x)$  are non-zero only where the edge state lies under the superconducting leads. They account for  the Andreev coupling arising from the two-dimensional electron gas (2DEG)   wavefunctions reaching  up to touch the superconductor as they drift under the  electrodes.     (See figure \ref{FIG:Edge}.)
 In contrast to the usual proximity effect, the topological protection of the QH edge modes means that this  interaction  {\it cannot\/} open a gap ---  but it  may, for example,  convert a  charge-$(e)$ right-going  spin-up electron into a charge-$(-e)$  right-going  spin-up hole, and in the process transfer a   spin-singlet pair of charge-$(e)$ electrons  from the Hall bar to the superconductor where they  merge with  the S-wave condensate.  
 
We have not included  Zeeman-energy term  to spilt the energy between the spin up and spin down edge modes. Such a  term adds only a multiple of the identity matrix to the BdG operator, and so has no effect on the subsequent analysis.
Further,  we assume that the energy scales of  relevance are  smaller than the energy gap of of the superconducting leads. We therefore regard  the parameters $|\Delta|$ as being externally imposed, and not to depend  the energy of the Hall-bar electrons, or on the temperature.

We can rewrite $H$ in the BdG form
\be
H= \int dx\left\{\left(\matrix{ \psid_{\uparrow}, \psi_{\downarrow}}\right)\left[\matrix{-iv_d (\partial _x-ieA) & |\Delta(x)| e^{i\theta(x)}\cr |\Delta(x)| e^{-i\theta(x)} &-iv_d (\partial_x+ieA)}\right] \left(\matrix{\psi_{\uparrow}\cr \psid_{\downarrow}}\right) \right\} +const.
\ee
Here  we have used an integration by parts together with  the anticommutation property  of the Fermi fields to write
\be
\int \{  \psid_{\downarrow}(- i v_d (\partial_x-ieA)) \psi_{\downarrow} \}dx = \int \{ \psi_{\downarrow}(- i v_d (\partial_x+ieA)) \psid_{\downarrow} \}dx +const.
\ee
This  rewriting is essentially a charge-conjugation  transformation that  makes manifest  the particle-hole symmetry of the linearized edge spectrum.  In particular, it  reveals that the charge-$(-e)$ spin-up holes created by $\psi_\downarrow$ move in the {\it same\/} direction as the charge-$(e)$  spin-up electrons created by $\psid_\uparrow$. 
The  ``constant''   contains the truly constant  ground-state energy of the spin down electrons, but also the term $- v_d e\int \delta(0) A(x)\,dx $ that  subtracts a background electric charge. This charge gets discarded as we switch to the charge-conjugate picture in which charge-$(-e)$  holes occupy the states that are not occupied by electrons. Keeping track of the ``constant'' restores  the  physical charge  when needed. 

The vector potential   $A$ acts as a chemical potential and controls the location of the Fermi energy.  In much of our discussion we will  assume that when  $\Delta=0$ the Fermi energy lies midway between  two  edge-mode energy levels.  This assumption is for illustrative purposes only. Indeed the detailed current/phase relation will depend sensitively on the exact location of the Fermi energy relative to the edge modes   because  varying $\theta$   can make a  level   cross the Fermi energy, change its  occupation, and cause a jump in the Josephson current.  The sensitivity  will  manifest itself   as Bohm-Aharonov oscillations  in the Josephson current  as a function of the  magnetic flux through the Hall bar \cite{ma}.

For our  mid-spaced $E_F$ we can make a gauge transformation to set $A\to 0$  at the expense of changing periodic boundary conditions to  antiperiodic ones, and simultaneously  redefining $\theta(x)$.  We  assume that we have done this.  
The Bogoliubov-deGennes (BdG) equation for the eigenmodes is therefore
\be
\left[-iv_d  \frac{\partial}{\partial x} +|\Delta(x)| e^{i\sigma_3\theta(x)}\sigma_1\right]\left(\matrix{u\cr v}\right) = E \left(\matrix{u\cr v}\right). 
\label{EQ:BdG}
\ee
Equation (\ref{EQ:BdG})  has a path-ordered exponential solution 
\be
\left(\matrix{u(x)\cr v(x)}\right) = e^{iEx/v_d}\,  {\rm P} \exp\left \{- i\int_0^x K(\xi)\, d\xi\right\} \left(\matrix{u(0)\cr v(0)}\right) ,
\ee
where $K(x)=   |\Delta(x)| e^{i\sigma_3\theta(x)}\sigma_1/v_d$ is a hermitian matrix. Note that, in distinction to  the usual BdG case, we did  {\it not\/} double the number of degrees of freedom when we constructed the BdG operator, so {\it all\/}  the BdG eigenmodes are needed.

Only a part  $\Omega$ (the union of the two regions under the SC electrodes)  of the perimeter of the Hall bar is in contact with  the superconductor, and we set    
\be
U= {\rm  P} \exp\left\{- i\int_\Omega K(\xi)\, d\xi\right\} \in  {\rm SU}(2).
\ee
As the  perimeter of the Hall bar forms a closed loop,  it was  reasonable to impose   periodic boundary conditions, but recall that  these were changed to antiperiodic boundary conditions by the gauge transformation that removed $A(x)$.   The eigenmodes of the BdG operator Hamiltonian are therefore determined from  the eigenvalues of $U$ by requiring that
\be
\left(\matrix{u_n\cr v_n}\right)= -e^{iE_nL/v_d} U\left(\matrix{u_n\cr v_n}\right).
\ee
Here $L$ is the length of the Hall-bar perimeter. Now the eigenvalues of $U$ will be of the form $e^{\pm i\phi}$ and so the energy eigenvalues are given by the requirement that $(E_n L/v_d)\pm \phi =\pi(2n+1)$, or 
\be
E_n=\frac {v_d}{ L} \left( \pi (2n+1) \mp \phi\right).
\ee
Note that  if $(u,v)^T$ is an eigenvector  of $U$ with eigenvalue $e^{i\phi}$ then $ -i\sigma_2(u^*,v^*)= ( -v^*, u^*)$   is an eigenvector  of $U$ with eigenvalue $e^{-i\phi}$. Consequently if   
$(u_n(x),v_n(x) )^T$ is an eigenfunction  of the BdG operator  corresponding to eigenvalue $E_n$, then 
$(-v^*_n(x),u^*_n(x) )^T$ is an eigenfunction  corresponding to energy $-E_n$. These facts  follow from 
\be
(i\sigma_2) \sigma_i (-i\sigma_2) =-\sigma_i^* \quad \Longrightarrow\quad (i\sigma_2) U^* (-i\sigma_2)= U,
\ee
and   give rise  to the  usual antilinear  S-wave BdG particle-hole symmetry ``$C$''  with $C^2=-{\rm Id}$. This symmetry  must be distinguished from the approximate particle-hole symmetry arising from our linearization of the quantum Hall edge-mode spectrum.

\begin{figure}
\includegraphics[width=3.0in]{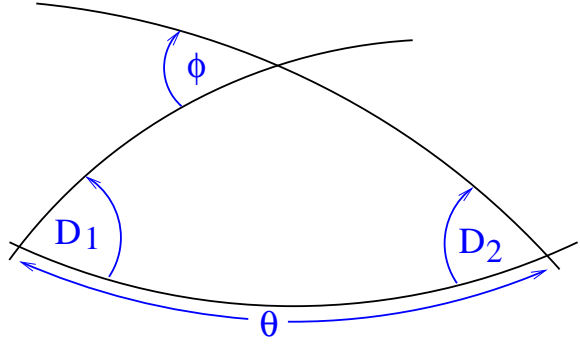}
\caption{\sl The spherical triangle that  relates   the eigen-phase  $\phi$ to   the order-parameter phase difference $\theta=\theta_2-\theta_1$.  }
\label{FIG:cosine}
\end{figure}

If the phase of the order parameter is constant in segments $\Omega_{1,2}$ (the superconducting leads) then 
$
U=U_2U_1$ where
\be
U_a= \left[\matrix{\cos D_a & -ie^{i\theta_a}\sin D_a\cr -ie^{-i\theta_a} \sin D_a &\cos D_a}\right], \quad a=1,2.
\ee
Here $D_a= |\Delta| w_a/v_d$ where $w_a$ is the width of lead $a$.
The eigenvalues of $U$ are $e^{\pm i\phi}$, and, by taking the trace of $U$, we see that  $\phi$ is given  by the spherical cosine rule:
\be
\cos \phi= \cos D_1 \cos D_2 - \cos \theta \sin D_1 \sin D_2.
\label{EQ:spherical}
\ee
The spherical triangle (see figure \ref{FIG:cosine}) arises because the matrices $U_1$ and $U_2$ are the spinor representations of successive ${\rm SO}(3)$ rotations  through angles $2D_1$ and $2D_2$  about  axes  separated  by the  angle $\theta$. It is shown in \cite{misner} that such  rotations can be combined through the use of mirrors that form the geodesic sides of the triangle.

From now on we  understand by ``$\phi$'', the  solution of  (\ref{EQ:spherical})   that lies in the range $0 \le\phi\le \pi$, and by the vector $(u,v)^T$ the corresponding eigenvector of $U$. We similarly  take ``$E_n$''  to mean  the combination
\be
E_n=  \frac {v_d} L \left( 2\pi (n+1/2) -\phi\right).  
\ee

Now we make the Bogoliubov transformation 
\be
 \left(\matrix{\psi_{\uparrow}(x)\cr \psid_{\downarrow}(x)}\right)= \sum_{n=-\infty}^\infty  \left\{ b_{n\uparrow}  \left(\matrix{u_n(x)\cr v_n(x)}\right)+ b^\dagger_{n\downarrow}
  \left(\matrix{-v_n^*(x)\cr \phantom- u_n^*(x)}\right)\right\}.
  \label{EQ:bogoliubov}
\ee
In order not to over-count,  we ensure that  the  modes are those that, after passing the  superconductor, take the form $(u_n(x), v_n(x))= e^{i(E_n x+\phi) }(u,v)$, and $(-v^*_n(x), u^*_n(x)) =e^{-i(E_nx/v_d+\phi)}(-v^*,u^*)$. 
The Fermionic anticommutation relations coupled with  the BdG eigenfunction  completeness relations then require that
\be
\{b_{n\downarrow}, b_{m\downarrow}\} = \{b_{n\uparrow}, b_{m\uparrow}\} =\{b_{n\downarrow}, b_{m\uparrow}\} =\{b^\dagger_{n\downarrow}, b^\dagger_{m\uparrow}\} = 0,
\ee
and 
\be
\{b^\dagger_{n\downarrow}, b_{m\downarrow}\} =   \{b^\dagger_{n\uparrow}, b_{m\uparrow}\} =\delta_{nm}.
\ee 

The Bogoliubov transformation simplifies  $H$ to
\be
H= \sum_{n=-\infty}^{\infty}  E_n(b^\dagger_{n\uparrow}b_{n\uparrow}- b_{n\downarrow}b^\dagger_{n\downarrow})+const,
\ee
the ``constant'' being the same one that was introduced earlier.  It is not really a constant as it depends on the gauge field $A$, but it is independent of $\theta(x)$.  Recall that the  $A$ dependence accounts for the total charge of the spin-down Fermi sea that was discarded when we made the particle-hole interchange for this spin component.     The minimum-energy  state is defined by the properties
\bea
b_{n\uparrow} \ket{0} =0, &\quad& E_n>0,\nonumber\\
b^\dagger_{n\uparrow}\ket{0}=0, &\quad& E_n<0,\nonumber\\
b_{n\downarrow} \ket{0} =0, &\quad& E_n>0,\nonumber\\
b^\dagger_{n\downarrow}\ket{0}=0, &\quad& E_n<0.
\eea
Using these, we compute the ground state energy to be
\be
E_{\rm ground}= \sum_{E_n<0}E_n -\sum_{E_n>0} E_n. 
\label{EQ:unregground}
\ee
The quantity  $E_{\rm ground}$ is formally divergent, but the physics resides entirely in the variation of $E_{\rm ground}$ with the phase difference  $ \theta = \theta_2-\theta_1$.  Now as we vary $\theta$  all  $E_n$ move in the same direction. The energy dependence on $\theta$ largely cancels between the two sums.  In order to  extract the   small,  but  non-zero, residuum we will have to regulate the sums in a controlled manner. This we do in the next section. 

\section{Computing the current}

\begin{figure}
\includegraphics[width=3.5in]{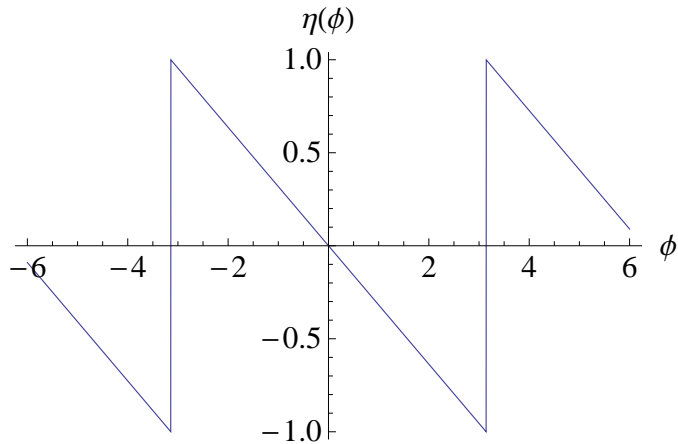}
\caption{\sl A plot of  $\eta(\phi)$ showing the $2\pi$ periodicity.}
\label{FIG:Etaregplot}
\end{figure}

Given a  Dirac-like spectrum of energy levels $-\infty< E_n<\infty $,  the associated ground-state charge and current can often be expressed  in terms of the spectral asymmetry \cite{bagmodel}.  This quantity   is defined \cite{atiyah,niemi} to be the regulated sum 
\be
\eta= \lim_{s\to 0}\left\{-\sum_{n=-\infty}^\infty {\rm sgn}(E_n) e^{-s|E_n|}\right\}.
\ee
For energies of our   form, $E_n= \alpha (2\pi (n+1/2)-\phi)$,  a  direct calculation shows that for $ -\pi< \phi<\pi$, we have 
\be
\left\{-\sum_{n=-\infty}^\infty {\rm sgn}( E_n) e^{-s|E_n|}\right\} = -\frac{\phi}{\pi}- \frac 1{6\pi}\left( \phi^3-\phi \pi^2\right)(\alpha s)^2+ O(s^4).
\ee
Thus 
\be
\eta(\phi)= - \frac{\phi}{\pi}, \quad -\pi <\phi<\pi,
\ee
and extends  with $2\pi$ periodicity in $\phi$, (see figure \ref{FIG:Etaregplot}).

\begin{figure}
\includegraphics[width=3.5in]{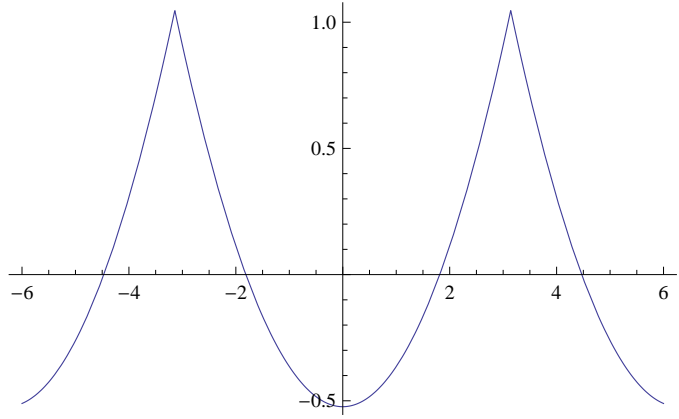}
\caption{\sl A plot  of $\alpha^{-1}(E_{\rm ground}(\phi))_{\rm reg}$ showing the $2\pi$ periodicity}
\label{FIG:Eregplot}
\end{figure}

We may similarly define  and compute an analytically-regulated version of the ground-state energy (\ref{EQ:unregground}): 
$$
(E_{\rm ground})_{\rm reg} =\lim_{s\to 0}  \left\{-\sum_{n=-\infty}^\infty {\rm sgn}(E_n) E_n e^{-s|E_n|}+\frac{1}{\pi \alpha s^2}\right\}=\alpha\left(
\frac{\phi^2}{2\pi}- \frac {\pi}{6}\right),\quad  -\pi <\phi<\pi.
$$
This quantity  also extends periodically outside the  range $-\pi <\phi<\pi$ --- see figure \ref{FIG:Eregplot}.
The subtraction  needed for the existence of the limit is independent of $\phi$, and the constant $-\alpha \pi/6$  is the same as would be obtained by zeta-function regularization 
\cite{difrancesco}. 
 Let us also compute 
\bea
\left(\frac{d  E_{\rm ground}}{d\phi}\right)_{\rm reg}&\stackrel{\rm def} {=}&
\lim_{s\to 0}\left\{-\sum_{n=-\infty}^\infty {\rm sgn}(E_n)\left(\frac{d E_n}{d\phi } \right) e^{-s|E_n|}\right\}\nonumber\\
&=&\lim_{s\to 0}\left\{\alpha\sum_{n=-\infty}^\infty {\rm sgn}(E_n) e^{-s|E_n|}\right\}\nonumber\\
&=&
\alpha \frac{\phi}{\pi},\nonumber
\label{EQ:star}
\eea
and observe that the regulated energy possesses  the comforting property that
\be
\frac{d}{d\phi}(E_{\rm ground})_{\rm reg}= \left(\frac{d E_{\rm ground}}{d\phi}\right)_{\rm reg}.
\ee
We will  relate these energy derivatives to  the  ground-state expectation value of the divergence of the current operator.

The current operator is
\be
j(x){=}-\frac{\delta H}{\delta A(x)}.
\ee
 If we include the contribution from the $A$ dependent ``constant'' when taking  the functional derivative, then the ground state current  is 
\bea
\vev{j(x)}&=& e v_d\eval{0}{\psid_\uparrow(x) \psi_\uparrow(x) +\psid_\downarrow(x) \psi_\downarrow(x)}{0}\nonumber\\
&=& 2ev_d \left(\sum_{E_n<0} |u_n(x)|^2+\sum_{E_n>0} |v_n(x)|^2\right).
\eea
If we ignore the ``constant,'' the current  becomes
\bea
\vev{j(x)}&=& ev_d\eval{0}{\psid_\uparrow(x) \psi_\uparrow(x) -\psi_\downarrow(x) \psid_\downarrow(x)}{0}\nonumber\\
&=& ev_d \sum_{E_n<0} \left( |u_n(x)|^2- |v_n(x)|^2\right)-ev_d \sum_{E_n>0} \left (|u_n(x)|^2- |v_n(x)|^2\right).  
\label{EQ:starstar}
\eea
These two currents  differ only by the subtraction of  $\sum_{n} (|u_n(x)|^2+|v_n(x)|^2)$ in the second case. This divergent  sum is ``$\delta(0)$'' and  independent of $x$  by  eigenvector completeness.  When it comes to computing the current flowing in and out at the leads we can use either expression therefore. The second expression is the most convenient, and so 
we    define
\be
\vev{j(x)}_{\rm reg}= \lim_{s\to 0}  \left\{-ev_d \sum_{n=-\infty}^\infty {\rm sgn}(E_n) \left (|u_n(x)|^2- |v_n(x)|^2\right) e^{-s|E_n|}\right\}. 
\ee
In our simple model $|u_n|^2(x)$ and  $|v_n|^2(x)$ are independent of $n$, but  {\it do\/}  depend on whether $x$ lies  between the superconducting leads or not.  This means that the edge-current differs in the two regions, and the difference  is due to  the Josephson current flowing in and out {\it via\/} the SC leads. We could compute  $|u_n|^2$ and  $|v_n|^2$  in the two  regions by diagonalizing the  matrix $U$, but it  is simpler, and more revealing,  to relate  the difference in the currents  to the variation of  the ground state energy with $\theta$. 

To do this we observe that
\bea
\left[\matrix{e^{i\chi/2}&0\cr 0 & e^{-i\chi/2}}\right] \left[\matrix{-iv_d (\partial _x-ieA) & |\Delta| e^{i\theta}\cr |\Delta| e^{-i\theta} &-iv_d (\partial_x+ieA)}\right] \left[\matrix{e^{-i\chi/2}&0\cr 0 & e^{+i\chi/2}}\right]\phantom{sin x}&&\nonumber\\
=\left[\matrix{-iv_d (\partial _x-i(eA+\chi'/2)) &| \Delta| e^{i(\theta +\chi)}\cr |\Delta| e^{-i(\theta+\chi)} &-iv_d (\partial_x+i(eA+\chi'/2))}\right].&&\nonumber
\eea
As the   similarity transformation does not change the eigenvalues of the BdG operator, we see that
\be
E_n[\theta, A] = E_n[\theta+\chi, eA+\chi'/2].
\ee
The effect on the energy eigenvalue of changing $\theta(x) \to \theta(x) +\delta\theta(x)$ is therefore identical to changing $eA\to eA-(\delta \theta)'/2$. By first-order perturbation theory we compute the latter effect to give
\bea
\delta E_n &=&\phantom- \eval{n}{\delta H}{n}\nonumber\\ 
&=&-  v_d  \int dx  \left(|u_n(x)|^2-|v_n(x)|^2\right) \delta A\nonumber\\
&=&\phantom-\frac 12v_d  \int dx  (|u_n(x)|^2-|v_n(x)|^2)\frac{\partial}{\partial x} \delta\theta(x)\nonumber\\
&=&- \frac 12 v_d\int dx \left\{ \frac{\partial}{\partial x} \left(|u_n(x)|^2-|v_n(x)|^2\right)\right\} \delta\theta(x).
\eea
Now, on  combining this last result with equations  (\ref{EQ:star}) and (\ref{EQ:starstar}), we find that
\bea
\delta (E_{{\rm ground}})_{\rm reg}&=&-\frac 1{2e}  \int \vev{\nabla \cdot j}_{\rm reg} \delta \theta(x)\, dx\nonumber\\
&=& \frac 1{2e} I_{\rm Josephson}(\delta\theta_2-\delta \theta_1).
\eea
Thus we see that   the general result 
\be
I_{\rm Josephson}=\left(\frac{2e}{\hbar}\right) \frac{d E_{\rm ground}}{d \theta}
\ee
is consistent with our regularization scheme.

\begin{figure}
\includegraphics[width=3.5in]{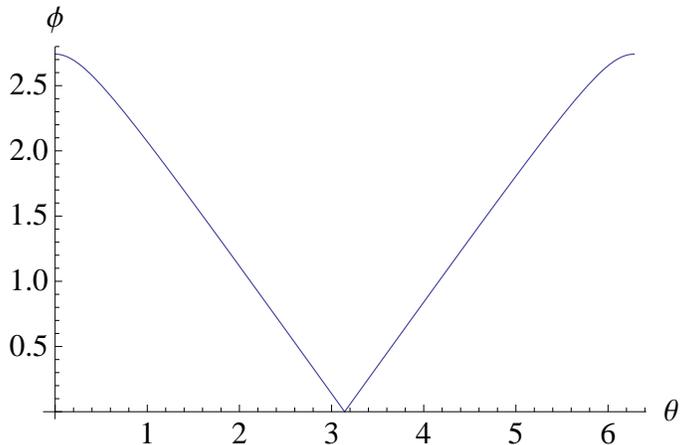}
\caption{\sl A plot of the eigen-phase   $\phi$ against $\theta$ for the case $D_1=D_2= \pi/2 -.2$. We are enforcing the condition $0\le \phi\le  \pi$ that is required by our   Bogoliubov transformation.}
\label{FIG:Phi1}
\end{figure}

\begin{figure}
\includegraphics[width=3.5in]{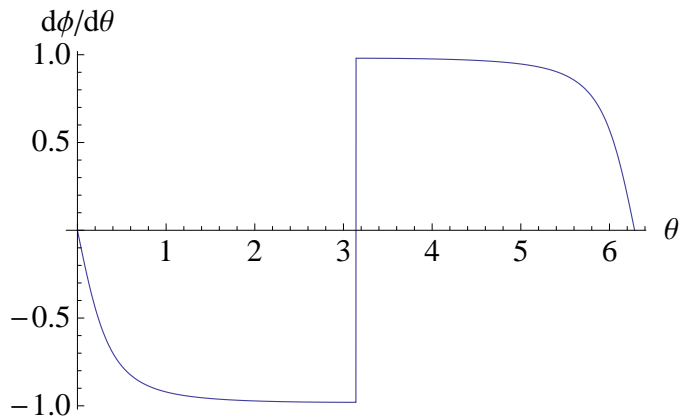}
\caption{\sl A plot of $d\phi/d\theta$ for  $D_1=D_2= \pi/2 -.2$}
\label{FIG:Dphi1}
\end{figure}

\begin{figure}
\includegraphics[width=3.9in]{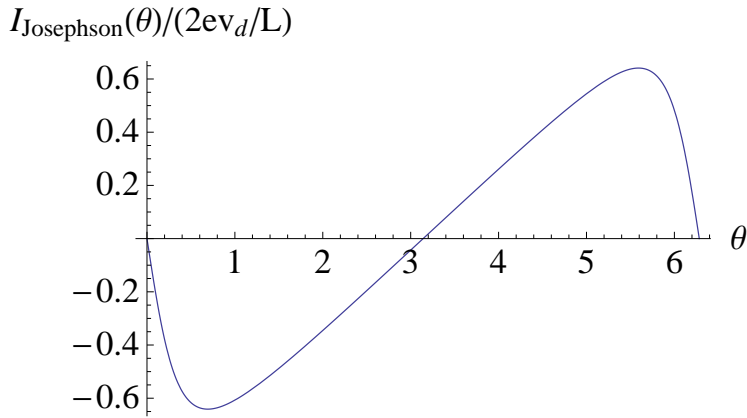}
\caption{\sl A plot of $I_{\rm Josephson}/(2ev_d/L) $ against $\theta$ for $D_1=D_2= \pi/2 -.2$. Observe how the discontinuities  combine to give a smooth result. As $D_{1,2}$ approach ``perfect coupling'' at $D_1=D_2=\pi/2$, the drops at $\theta=0, 2\pi$ steepen, and become  level-crossing discontinuities.  }
\label{FIG:Josephson1}
\end{figure}

From  
\be
E_{\rm ground} = \frac{v_d}{L}\left( \frac{\phi^2}{2\pi}-\frac \pi 6 \right), \quad 0  \le \phi\le \pi.
\ee
we have
\be
I_{\rm Josephson}= 2e \frac{d}{d\theta}(E_{\rm ground})_{\rm reg}= 2e\frac{d}{d\phi}(E_{\rm ground})_{\rm reg}\frac{d\phi}{d\theta}
\ee
 Figures \ref{FIG:Phi1},\ref{FIG:Dphi1},\ref{FIG:Josephson1} show how theses  ingredients assemble to give the   current/phase relation.

To gain further  insight, consider the case of ``perfect  coupling,'' where $\sin D_a=1$ and $\phi = \pm(\theta_2-\theta_1+\pi)$. In this case 
\bea
U&=&  \left[\matrix{0&- ie^{i\theta_2}\cr- ie^{-i\theta_2}  &0 }\right]\left[\matrix{0&- ie^{i\theta_1}\cr-ie^{-i\theta_1}  &0 }\right]\nonumber\\
&=&\left[\matrix {- e^{i(\theta_2-\theta_1)} &0 \cr 0& -e^{-i(\theta_2-\theta_1)}}\right],
\eea
and so  $\phi=(\theta_2-\theta_1)+\pi$. In the absence of relaxation, each $2\pi$ turn of $\theta$ would put  another particle into both the spin-up and spin down sea. In equilibrium however, the state  ceases  to be occupied as soon at its energy becomes positive. This  change in occupation  leads to a jump in the Josephson current as the state crosses  the Fermi energy and its contribution is lost. The maximum possible current occurs just before or after the jump,  and has $I_{\rm max} =\pm  2e v_d/L$.  For $v_d\sim 10^6\, {\rm m/s}$ and a perimeter of about $1\, {\rm mm}$ we get 
an upper bound on  the Josephson current of  about $1\, {\rm nA}$. This is consistent with the estimate of Ma and Zyuzin \cite{ma}. 

A physical picture for this upper bound is as follows:  At the phase difference corresponding to the ``jump,'' we have a spin-up/spin-down pair of    levels lying exactly at the Fermi energy.  At perfect coupling, the extreme equilibrium currents correspond to  two possible cases:  i) between the leads both  zero-energy levels are  empty whilst  outside they are occupied, ii) between the leads both zero-energy  levels  are occupied and outside they are empty.  Levels in the Dirac sea   that are {\it not\/}  at the Fermi energy cannot  be left empty by a passage under a lead, as this would lead to the energy being different in different regions and this is  not possible in an energy eigenstate.  Only the topmost energy level can contribute to the equilibrium Josephson current therefore, and this is the reason why the Josephson current is so small. To estimate its magnitude we note that  
in case (i), in each passage round the perimeter of the Hall bar, a  pair of electrons passes from the Hall bar  to the first lead and is returned to the Hall bar from the second lead. In case (ii) in each orbit a pair of electron passes from the first lead to the Hall bar, and is collected from the Hall bar at the second lead.  This physical picture shows that the  two possible Josephson currents are equal and opposite and have magnitude $|I_{\rm max}| = 2e v_d/L$.  (Because it is easy to get confused by Bogoliubov transformations,  we provide,   in Appendix A,  a  more detailed  description of   what happens to the particle content of the many-body eigenstates as they pass under the superconducting leads.)

\section{Comparison with  perturbation theory}

The analytic regularization method used in the computations in the previous sections is standard in relativistic field theory \cite{bagmodel}, but is perhaps less familiar in  superconducting  applications. As a check on its validity it is worthwhile (and non-trivial) to  compare our all-orders in $D_1$ and  $D_2$ calculation with  conventional  perturbation theory.  

In the weak-coupling regime, where $D_1$ and $D_2$ are small, the spherical cosine rule reduces to  
\be
\phi^2 =  D_1^2+D_2^2 +2 D_1D_2 \cos \theta +O(D^3).
\ee   In this limit the   ground-state energy and zero-temperature and  Josephson current become  
\be
 E_{\rm ground}(\theta)= \frac{v_d}{L} \frac 1 {2\pi}(D_1^2+D_2^2+  2D_1D_2 \cos\theta),
 \label{EQ:smalld}
\ee
and  hence
\be
I_{\rm Josephson}=- \frac{2e v_d}{\pi L}   D_1D_2 \sin \theta. 
\ee
We begin by verifying that (\ref{EQ:smalld}) is  correctly reproduced by the perturbation expansion.

The Euclidean chiral propagator  for zero temperature and anti-periodic spatial boundary conditions   is 
\be
\eval{0}{T\psid_a(z_1)\psi_b(z_2)}{0}=\delta_{ab}G(z_1-z_2)=\frac 1 {2iL}\frac{\delta_{ab}}{\sin[\pi (z_1-z_2)/L]}
\ee
where $a,b= \uparrow,\downarrow$ and $ z=x+iv_d\tau$.
The change in the $\Delta=0$  ground-state energy due to the interaction
\be
H_{\rm int} =\int |\Delta(x)|\left(e^{i\theta(x)}\psid_\uparrow(x)\psid_\downarrow(x)  + e^{-i\theta(x)}\psi_\downarrow(x) \psi_\uparrow(x)\right)dx
\ee
 occurs at second order in $|\Delta|$, and is
\be
\delta E_{\rm ground} =-\int dx_1 \int dx_2 \int_{-\infty}^{\infty}d\tau |\Delta(x_1)| |\Delta(x_2)| e^{\theta(x_1)}e^{-i\theta(x_2)} \eval{0}{T  \psid_\uparrow(z_1)\psid_{\downarrow}(z_1)\psi_{\downarrow}(z_2)\psi_{\uparrow}(z_2)}{0}.
\ee
Here $\tau=\tau_2-\tau_1$ is the Euclidean time interval between $z_2$ and $z_1$.
Now 
\be
\eval{0}{T  \psid_\uparrow(z_1)\psid_{\downarrow}(z_1)\psi_{\downarrow}(z_2)\psi_{\uparrow}(z_2)}{0}= [G(z_1-z_2)]^2
\ee
by Wick's theorem, and
\be
\frac 1{4L^2} \int_{-\infty}^{\infty} \frac{1}{(\sin[\pi (x_1-x_2+iv_d\tau)/L])^2}\, d\tau=\left(\frac{1}{2 \pi L v_d}\right)
\ee
is independent of the separation  $x_1-x_2$ unless $x_1-x_2=0$ (mod $L$). The perturbation integral has four contributing regions: i) both $x_1$ and $x_2$ in lead 1, ii) both $x_1$ and $x_2$ in lead 2, iii)  $x_1$ in lead 1, $x_2$ in lead 2, iv)  $x_1$ in lead 2, $x_2$ in lead 1. Recalling that $D_a= |\Delta|w_a/v_d$, these combine to give
\bea
\delta E_{\rm ground} &=& v_d^2 (D_1^2+D_2^2+2D_1D_2 \cos\theta) \frac{1}{2\pi Lv_d}\nonumber\\
&=&\frac{v_d}{2\pi L} (D_1^2+D_2^2+2D_1D_2 \cos\theta). 
\eea
This expression coincides with the weak coupling limit of the all-orders calculation.

We can extend the comparison to non-zero  temperature.
At  temperature $T=\beta^{-1}$,  the Josephson current can be written as 
\be
I_{\rm Josephson}=\left(\frac{2e}{\hbar}\right) \frac{d F}{d \theta}
\ee
where $F$ is the free energy. 
For a general spectral shift $\phi$, we use standard methods to write down the  partition function 
\bea
Z&=&\exp\left\{ -\beta F[\phi, \beta]\right\}\nonumber\\
 &=&\exp\left\{-\frac{\beta v_d}{L} \left(\frac{\phi^2}{2\pi}-\frac{\pi}{6}\right)\right\}\prod_{N=1}^{\infty} (1+wq^{2n-1})^2(1+w^{-1}q^{2n-1})^2\nonumber\\ 
 &=&(\eta(q))^{-2}\left[ \sum_{n=-\infty}^{\infty} \exp\left\{ - \frac{ v_d \beta}{2\pi L}\frac 12 \left( 2 \pi n+\phi\right)^2\right\}\right]^2,
\eea
where $q= \exp\{-\pi \beta v_d/L\}$, $w= \exp\{-\beta v_d\phi/L\}$,
and
$$
\eta(q)=q^{1/12}\prod_{n=1}^\infty(1-q^{2n})
$$
is the Dedekind eta function. We used the Jacobi triple-product formula to pass from the second line to the third.
 The sum in the expression for $Z$ is  squared because there are two independent Fermi seas (spin up and spin down) and their contributions to the partition function are symmetric under the interchange of $\phi$ with $-\phi$. 
 By using the Poisson summation formula, we  can rewrite the partition function  as 
 \bea
\exp\left\{ -\beta F[\phi, \beta]\right\} &=& (\eta(q))^{-2}\frac{L}{v_d \beta}  \left[\sum_{n=-\infty}^\infty \exp\left\{ - \frac 12 \frac{2\pi L}{ v_d\beta} n^2 +in\phi\right\}\right]^2 \nonumber\\
 &=&  (\eta(q))^{-2}\frac{L}{v_d \beta} \left[ \theta_3(\phi/2\pi|iL/v_d\beta)\right]^2
 \eea
 Thus the free energy is given by 
 \be
 F[\phi, \beta]  =c-\frac 2 \beta \ln   \theta_3(\phi/2\pi|iL/v_d\beta).
 \ee
 where $c$ does not depend on $\phi$. For small spectral shifts $\phi$, we can Taylor expand 
 \be
  F[\phi, \beta]  =c -\frac 1 {\beta} \phi^2 \frac{d^2}{d\phi^2}  \ln   \theta_3(\phi/2\pi|L/v_d\beta)  +O(\phi^4).
  \label{EQ:smallF}
  \ee

\begin{figure}
\includegraphics[width=4.0in]{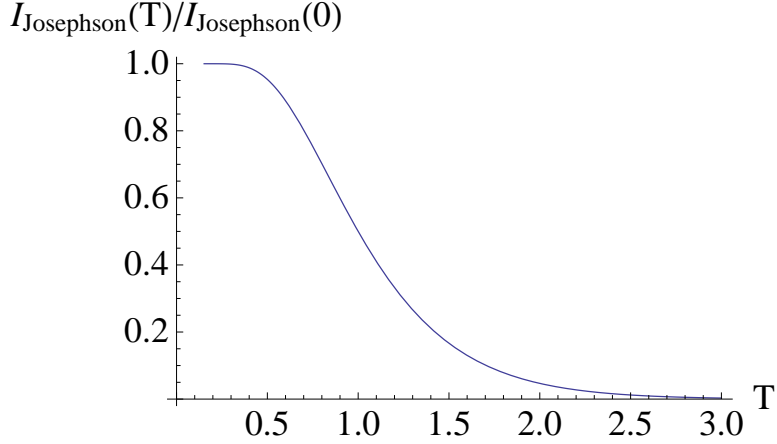}
\caption{\sl A plot of the effect of temperature on the perturbative  Josephson current.. The horizontal axis is temperature in units of $\hbar v_d/L$. We see an effect as soon as the temperature becomes comparable with the $2\pi \hbar v_d/L$ level  spacing of the edge energy states. }
\label{FIG:temperature}
\end{figure}

We would now like to compare the expression  (\ref{EQ:smallF})  with that obtained by perturbation theory.
At  finite temperature the chiral propagator   becomes 
\be
\eval{0}{T\psid_a(z)\psi_b(0)}{0}\to G(z)=\frac 1 {2\pi iL}\frac{\theta'(0|iv_d\beta/L)}{\theta(z/L|iv_d\beta/L)}\frac{\theta_3(z/L|iv_d\beta/L)}{\theta_3
(0|iv_d\beta/L)}.
\label{EQ:matsubara}
\ee
Here we are using the theta function definitions from \cite{chandra}, in which 
\bea
\theta(z|\tau) &=& \sum_{m=-\infty}^{\infty} \exp\left\{ i\pi \tau(m+1/2)^2+2\pi i (m+1/2)(z+1/2)\right\}, \nonumber\\
\theta_3(z|\tau) &=& \sum_{m=-\infty}^{\infty} \exp\left\{ i\pi \tau m^2+2\pi i mz\right\}.
\eea
Thus  $\theta(z|\tau)$ is odd under $z\leftrightarrow -z$, while  $\theta_3(z|\tau)$ is even.
These  properties  were the ingredients   used to   assemble 
(\ref{EQ:matsubara}), which  is  specified uniquely by requiring the propagator to be  analytic,  doubly anti-periodic  
\be
G(z+L)=-G(z), \quad G(z+iv_d\beta)=-G(z),
\ee
and for small $z$ to obey
\be
G(z) \sim \frac{1}{2\pi i}\frac 1 z.
\ee
It is this last property that makes it a Green function.

In terms of $G(z)$ we now have
\be
\delta E_{\rm ground} =-\int dx_1 \int dx_2 \int_{0}^{\beta}d\tau |\Delta(x_1)| |\Delta(x_2)| e^{\theta(x_1)}e^{-i\theta(x_2)} [G(x_1-x_2+iv_d \tau)]^2.
\ee
The $x_a$ integrals are the same as before, and, 
although it is little more complicated, the  integral over $\tau$ can still  be evaluated in closed form. 
We begin by  observing   that $[2\pi i G(z)]^2$ is analytic, has a double pole $1/z^2$ at the origin, is doubly periodic with periods $\omega_1=L$ and $\omega_2=iv_d \beta$, and (from the $\theta_3(z|\tau)$   in the numerator) has a double zero at $z=\frac 12(\omega_1+\omega_2)$.  These properties  are  sufficient  to show    that  
\be
 [2\pi i G(z)]^2 = \wp(z\,|\,\omega_1,\omega_2)-e_3,
 \ee
 where $\wp(z\,|\,\omega_1,\omega_2)$ is the Weierstrass elliptic function, and
 \be
 e_3\equiv \wp(\{\omega_1+\omega_2\}/2\,|\,\omega_1,\omega_1).
 \ee
 The Weierstrass zeta function is defined so that 
 \be
 \frac{d}{dz} \zeta(z\,|\,\omega_1,\omega_2) = - \wp(z\,|\,\omega_1,\omega_2),  
 \ee
 together with initial condition 
 \be
 \lim_{z\to 0}\left \{\zeta(z)-\frac 1 z\right\}=0.
 \ee
 We may therefore evaluate  the $\tau$ integral in terms of tabulated functions:
\bea
 \int_a^{a+\omega_2} [2\pi i G(z)]^2 dz&=& -\zeta(a+\omega_2)+\zeta(a)- \omega_2e_3 \nonumber\\
 &=& -2\eta_2- \omega_2e_3,\nonumber\\
 &=&\left. \frac 1{\omega_2}  \frac{d^2}{dz^2} \ln \theta_3(z|-\omega_1/\omega_2)\right|_{z=0}.
 \label{EQ:zetaomega}
\eea
Here $2\eta_2\equiv  \zeta(a+\omega_2)-\zeta(a)=2 \zeta(\omega_2/2)$ is independent of $a$.
The quantities  in the second line of (\ref{EQ:zetaomega}) are   available in Mathematica\texttrademark, 
and we use them to    plot $I_{\rm Josephson}(T)/I_{\rm Josephson}(0)$ in \hbox{Figure \ref{FIG:temperature}.}  

It takes a little more work to obtain the  logarithmic derivative appearing in the last line of (\ref{EQ:zetaomega}),  and so we relegate its derivation to Appendix B. 
Accepting that the claim  is correct, and putting in the dimensionful constants,  we confirm  that  our all-orders evaluation of the free energy coincides with the perturbation theory calculation in the weak coupling regime .

\section{Discussion}

We have shown that the maximum possible Josephson current for a pair of spin-up/spin-down QH edge states is rather small for typical Hall bar geometries.   The bound is small because the relevant length and energy scales are set by the perimeter of the Hall device rather than the separation of the  superconducting probes. Also, unlike a typical Josephson device, there is only one conduction channel per pair of edge modes.  This last observation means that nothing is to be gained by making the superconducting leads overlay deeper into the Hall bar. 
   
   It may seem strange that we have so far discussed quantum Hall physics with no mention of the magnetic field that is necessary for its existence. The field, however,  has only a few consequences for our discussion.  Obviously  the superconducting leads must be constructed of materials that remain superconducting in a field of few Tesla at temperatures of about 1K, but this is not hard to achieve.
The leads must also be  narrow  enough that the order-parameter phase does not vary widely within the part of the lead that is actively coupled to the 2DEG. A subtle point in this regard affects the claim that the Josephson current is independent of the separation of the leads.  The phase difference $\theta$ that we have equated to  $\theta_2-\theta_1$ should   be understood as the gauge invariant quantity $\theta=\theta_2-\theta_1-2e\int_{x_1}^{x_2} A\,dx$. Now a  quantum of magnetic  flux lies between each of the edge-state energy levels and if the effective ``$\theta$''  is not to vary with the energy level index $n$, only a small fraction of this flux should pass between the leads. The leads should not be spaced apart by more than a small fraction of the perimeter. 

An  effect that we have not considered here, and one that may well allow for larger currents,  is ``edge reconstruction''  \cite{beenakker,chang,chklovski}. A reconstructed edge, with its alternating strips of compressible and incompressible 2DEG can allow many more levels to lie exactly at the fermi energy and so have their occupation number changed without a change in energy.  These levels have zero drift velocity, however,  so it unlikely that they contribute significantly to the Josephson current.  
  
 \section{Acknowledgements}   We  thank Tony Leggett for interesting us the problem, and also Jim Eckstein and Stephanie Law  Toner for explaining their work on QHE superconductor interfaces. The contribution of MS to this  project was supported by the National Science Foundation  under grant  DMR 09-03291. The work of   YL was supported by the US Department of Energy, Division of Materials Sciences, under  award DE-FG02-07ER46453, admistered through the Frederick Seitz Materials Research Laboratory  at the University of Illinois.  
 
 \section{Appendix A}
 
 The maximum possible Josephson current occurs when we  have both perfect coupling  ($\sin D_1=\sin D_2=1$) and   $\cos \theta=1$.  In this special case we have 
 \be
 U_1=U_2= \left[\matrix{ 0 &-i \cr- i &0}\right], \quad U= U_2U_1= - \left[\matrix{ 1 &0 \cr 0 &1}\right].
 \ee 
 The  Bogoliubov mode-expansion  (\ref{EQ:bogoliubov}) then becomes 
 \be
 \left(\matrix{\psi_{\uparrow}(x)\cr \psid_{\downarrow}(x)}\right)= \sum_{n=-\infty}^{\infty} \left\{ b_{n\uparrow} \frac{1}{\sqrt{L}} \left(\matrix{1\cr 0}\right)e^{2\pi i nx/L}+ b^\dagger_{n\downarrow}
\frac{1}{\sqrt{L}}   \left(\matrix{0\cr 1}\right)e^{-2\pi i nx/L}\right\}
\ee
for $x$ in region (1), and
\be
 \left(\matrix{\psi_{\uparrow}(x)\cr \psid_{\downarrow}(x)}\right)= \sum_{n=-\infty}^{\infty} \left\{ b_{n\uparrow} \frac{1}{\sqrt{L}}  \left(\matrix{0\cr- i}\right)e^{2\pi i nx/L}+ b^\dagger_{n\downarrow}
 \frac{1}{\sqrt{L}}  \left(\matrix{-i\cr 0}\right)e^{-2\pi i nx/L}\right\}
\ee
for $x$ in region (2). (The numbering  of the regions refers to figure \ref{FIG:Hallbar}.)

In these mode-expansions, the operators $b_{n\uparrow} $ and $b^\dagger_{n\downarrow}$ annihilate or create {\it quasiparticles\/} with energy $|E_n|=2\pi v_d  |n|/L$.  We compare these expansions with   the free-particle plane wave expansion
\be
 \left(\matrix{\psi_{\uparrow}(x)\cr \psid_{\downarrow}(x)}\right)= \sum_{n=-\infty}^{\infty} \left\{ a_{n\uparrow} \frac{1}{\sqrt{L}}  \left(\matrix{1\cr 0}\right)e^{2\pi i nx/L}+ a^\dagger_{n\downarrow}
\frac{1}{\sqrt{L}}   \left(\matrix{0\cr 1}\right)e^{-2\pi i nx/L}\right\},
\ee 
where the operators $a_{n\uparrow}$ and $a^\dagger_{n\downarrow}$ annihilate and create {\it electrons}.  We see that we can identify 
\bea
b_{n\uparrow} = a_{n\uparrow},&&  b^\dagger_{n\uparrow} = a^\dagger_{n\uparrow}\nonumber\\
b_{n\downarrow} = a_{n\downarrow},&&  b^\dagger_{n\downarrow} = a^\dagger_{n\downarrow}
\eea
in region (1), and 
\bea
b_{n\uparrow} = ia^\dagger_{-n\downarrow},&&  b^\dagger_{n\uparrow} = -ia_{-n\downarrow}\nonumber\\
b_{n\downarrow} = -ia^\dagger_{n\downarrow},&&  b^\dagger_{n\downarrow} = +ia_{-n\uparrow}
\eea
in region (2).  We now use these identifications to examine  what happens to the particle content of the  many-body eigenstates as they drift under  the superconducting leads.

We first note a minimum-energy eigenstate  must be  annihilated by $b_{n\uparrow} $ and $b_{n\downarrow}$ for $n>0$, and by $b^\dagger_{n\uparrow}$ and $b^\dagger_{n\downarrow}$ for $n<0$.   Let us define the eigenstate  $\ket{0}$ by requiring that it is killed by all these operators, and also  by $b_{0\downarrow}$ and $b_{0\uparrow}$.
Then the  states  
\be
\ket{0}, \quad b^\dagger_{0\uparrow} \ket{0},\quad   b^\dagger_{0\downarrow} \ket{0},\quad 
  b^\dagger_{0\downarrow} b^\dagger_{0\uparrow} \ket{0},
\ee
all have the same energy, making the ground state four-fold degenerate.  

With the operator identifications established above, we find that
\be
\ket{0} = \prod_{n=-\infty}^{-1}  (a^\dagger_{n\downarrow}a^\dagger_{n\uparrow}) \ket{\rm empty} 
\ee
when  $x$ lies in region (1), 
but in region (2), where $b_{0\uparrow}$ and $b_{0\downarrow}$ are identified with $a^\dagger_{0\downarrow}$ and $a^{\dagger}_{0\uparrow}$ respectively, we must have
\be
\ket{0}\propto a^\dagger_{0\downarrow}a^\dagger_{0\uparrow}  \prod_{n=-\infty}^{-1} (a^\dagger_{n\downarrow}a^\dagger_{n\uparrow}) \ket{\rm empty},
\ee
for it still to be annihilated by $b_{0\uparrow}$ and $b_{0\downarrow}$.
We see that the occupation number of the energy levels  for $n<0$ are unchanged, but  $\ket{0}$    picks up  a pair of $n=0$  electrons  from the superconducting lead as it passes under it. Similarly the state $  b^\dagger_{0\downarrow} b^\dagger_{0\uparrow} \ket{0}$ {\it  loses\/} a pair  from the $n=0$ level.

The state $ b^\dagger_{0\uparrow} \ket{0}$ is annihilated by $a^\dagger_{0\uparrow}$ and $a_{0\downarrow}$ in region (1),  and these become  respectively $a_{0\downarrow}$ and $a^\dagger_{0\uparrow}$ in region (2). The particle content of this  state is  unaffected by its passage under the lead therefore. Similarly $ b^\dagger_{0\downarrow} \ket{0}$ retains its particle content.

Now consider an   excited state, for example  $b^\dagger_{m\uparrow}  b^\dagger_{0\uparrow} \ket{0}$ with $m>0$.  This state   has energy $E= 2\pi v_d m/L$. In region (1) it has particle content 
\be
a^\dagger_{m\uparrow}  a^\dagger_{0\uparrow}  \prod_{n=-\infty}^{-1}  (a^\dagger_{n\downarrow}a^\dagger_{n\uparrow}) \ket{\rm empty},  
\ee
and so consists of  a Dirac sea together with   an electron in  a positive energy level.
In region (2) it becomes 
\be
 a_{-m\downarrow}  a^\dagger_{0\uparrow}  \prod_{n=-\infty}^{-1}  (a^\dagger_{n\downarrow}a^\dagger_{n\uparrow}) \ket{\rm empty},
\ee
which consists of a  Dirac sea  which has lost an electron from a negative energy level.
After passing the superconductor therefore, the state  has the same energy and spin, but the electron has become a hole.

 \section{Appendix B}
 
We wish to establish the third line of (\ref{EQ:zetaomega}), which reads   
\be
\int_a^{a+\omega_2} (\wp(z|\omega_1,\omega_2)-e_3) dz=\left. \frac 1{\omega_2}  \frac{d^2}{dz^2} \ln \theta_3(z|-\omega_1/\omega_2)\right|_{z=0}.
\label{EQ:desired}
\ee
This result follows
indirectly from  the related integral
\bea
\int_a^{a+\omega_1}\left\{\wp(z|\omega_1,\omega_2)-e_3\right\}dz &=& -2\eta_1 -\omega_1 e_3\nonumber\\
&=& \frac 1{\omega_1} \frac{\theta''_3(0|\omega_2/\omega1)}{\theta_3(0|\omega_2/\omega_1)},\nonumber\\
&=&\left. \frac 1{\omega_1} \frac{d^2}{dz^2} \ln \theta_3(z|\omega_2/\omega_1)\right|_{z=0}. 
\label{EQ:logtheta}
\eea
Here  we require  ${\rm Im\,} (\omega_2/\omega_1)>0$ for the theta functions to converge.  To establish (\ref{EQ:logtheta}) we observe that   second line follows from the first   by   combining  
two  standard formul\ae:
\be
e_3 = \frac 1{\omega_1^2} \left\{\frac 13 \frac{\theta'''(0|\tau)}{\theta'(0|\tau)}- \frac{\theta''_3(0|\tau)}{\theta_3(0|\tau)}\right\},
\ee
(\cite{chandra} Eq 5.2),
and 
\be
2\eta_1 = - \frac 1{\omega_1}\frac 13  \frac{\theta'''(0|\tau)}{\theta'(0|\tau)},
\ee
(\cite{whittaker} \textsection 21.43.) Here $\tau=\omega_2/\omega_1$  with ${\rm Im\,}\tau>0$.  The  third line of (\ref{EQ:logtheta}) follows from the second because $\theta_3'(0|\tau)=0$.

To derive (\ref{EQ:desired}) however,  we need  the integral over the $\omega_2=iv_d\beta$  imaginary period,  and not over the  $\omega_1=L$  real  period.  Because of the positivity condition on the imaginary part of $\tau$, we cannot change the integration path  by   merely  interchanging  $\omega_1\leftrightarrow \omega_2$ in equation (\ref{EQ:logtheta}).   We need to be more subtle. By changing   $(\omega_1,\omega_2)\to (-\omega_2,\omega_1)$  in (\ref{EQ:logtheta}), we obtain 
\be
-\left. \frac 1{\omega_2} \frac{d^2}{dz^2} \ln \theta_3(z|-\omega_1/\omega_2)\right|_{z=0}=\int_a^{a-\omega_2}\left\{\wp(z|-\omega_2,\omega_1)-e_3\right\}dz.
\ee
This last equation  is legitimate because  ${\rm Im\,}(\omega_2/\omega_1)>0$ implies that ${\rm Im\,}(-\omega_1/\omega_2)>0$.
We now manipulate
\bea
RHS &=& -  \int_a^{a+\omega_2}\left\{\wp(z|-\omega_2,\omega_1)-e_3\right\}dz\nonumber\\
&=&  -  \int_a^{a+\omega_2}\left\{\wp(z|\,\omega_1,\omega_2)-e_3\right\}dz,
\eea
where the last line follows from the invariance of $\wp(z|\,\omega_1, \omega_2)$ under modular transformations
\be
\left(\matrix{\omega_1'\cr \omega_2'}\right)= \left(\matrix{ a& b\cr c& d}\right)\left(\matrix{\omega_1\cr \omega_2}\right), \qquad  \left(\matrix{ a& b\cr c& d}\right)\in {\rm SL}(2,{\mathbb Z}).
\ee
From this we immediately deduce eq.\ (\ref{EQ:desired}).

\end{document}